\documentclass[fleqn,usenatbib]{mnras}

\usepackage{newtxtext,newtxmath}
\usepackage[T1]{fontenc}
\usepackage{subfigure}
\usepackage{tablefootnote}
\DeclareRobustCommand{\VAN}[3]{#2}
\let\VANthebibliography\thebibliography
\def\thebibliography{\DeclareRobustCommand{\VAN}[3]{##3}\VANthebibliography}

\usepackage{graphicx}	% Including figure files
\usepackage{amsmath}	% Advanced maths commands

\usepackage{amssymb}
\usepackage{mathtools}

\title[Optical Variability of RL-NLSy1 galaxies]{Probing jet-induced optical variability across timescales in radio-loud NLSy1 galaxies}

\author[Vivek Kumar Jha et al.]{
Vivek Kumar Jha,$^{1}$\thanks{E-mail: vivekjha.aries@gmail.com (VKJ)}
Anshul Kumar Sharma,$^{2}$
Madhu Sudan,$^{2}$
and Hum Chand $^{2}$
\\
$^{1}$National Centre for Radio Astrophysics, Tata Institute of Fundamental Research, Post Bag 3, Ganeshkhind, Pune, 411007; India\\
$^{2}$Department of Physics and Astronomical Science, Central University of Himachal Pradesh, Dharamshala, 176215, India\\}

\date{Accepted XXX. Received YYY; in original form ZZZ}

\pubyear{\the\year{}}

\begin{document}
\label{firstpage}
\pagerange{\pageref{firstpage}--\pageref{lastpage}}
\maketitle


\begin{abstract}

We investigate optical variability across multiple timescales in a sample of radio-loud narrow-line Seyfert~1 (RL-NLSy1) galaxies, including $\gamma$-ray detected, $\gamma$-ray undetected, and non-jetted systems along with a comparison set of highly polarised core-dominated quasars (HPQs). Using Zwicky Transient Facility light curves, we measure fractional variability ($F_{\rm var}$) and rest-frame structure functions (SFs) to test whether short-term jet-linked variability is reflected in long-term behaviour. $\gamma$-ray detected RL-NLSy1s and HPQs show steeply rising SFs, revealing strong long-term coherence despite modest $F_{\rm var}$, consistent with Doppler-boosted synchrotron emission from relativistic jets. Non-jetted RL-NLSy1s exhibit the highest $F_{\rm var}$ but plateauing SFs, indicative of stochastic, disc-driven fluctuations lacking long-term coherence. $\gamma$-ray undetected RL-NLSy1s show the lowest $F_{\rm var}$ and nearly flat SFs, consistent with weak or absent jet activity across all timescales. Colour-magnitude trends show that jet-dominated sources exhibit redder-when-brighter behaviour, whereas disc-dominated systems exhibit bluer-when-brighter trends. These results show that SF-derived temporal coherence, rather than variability amplitude alone, is a promising diagnostic of jet dominance and orientation, offering a framework for interpreting AGN variability in forthcoming time-domain surveys.

\end{abstract}

\begin{keywords}
galaxies: active -- galaxies: jets -- galaxies: Seyfert -- galaxies: nuclei -- quasars: general --  techniques: photometric
\end{keywords}

%%%%%%%%%%%%%%%%%%%%%%%%%%%%%%%%%%%%%%%%%%%%%%%%%%

%%%%%%%%%%%%%%%%% BODY OF PAPER %%%%%%%%%%%%%%%%%%

\section{Introduction}

Variability is a fundamental feature of active galactic nuclei (AGN), observed across all wavelengths and timescales from minutes to years \citep{1997ARA&A..35..445U}. These fluctuations provide key insights into the physics of accretion discs and relativistic jets \citep{1985ApJ...298..114M,2019ARA&A..57..467B}. While ensemble AGN variability is often modelled stochastically \citep{2013ApJ...765..106Z}, most studies focus on either intra-day or long-term behaviour, rarely both in the same sources.

Radio-loud narrow-line Seyfert 1 (RL-NLSy1) galaxies present an intriguing subclass \citep{1989AJ.....98.1195K,2006AJ....132..531K,2008ApJ...685..801Y}. While optically they resemble classical NLSy1s \citep{1985ApJ...297..166O}, some exhibit compact flat-spectrum radio cores and $\gamma$-ray emission, pointing to relativistic jets \citep{2009ApJ...699..976A,2009ApJ...707L.142A,2015A&A...575A..13F,2020Univ....6..136F,2019Galax...7...87D}. These systems often lie in disk galaxies with lower black hole masses \citep{2015ApJS..221....3G}, and are considered low-mass, high-Eddington analogues of blazars \citep{refId0,2015A&A...575A..13F}.  Morphologically, they range from superluminal jets \citep{2011ApJ...738..126D, 2012MNRAS.426..317D, 2016AJ....152...12L} to absorbed or misaligned sources \citep{2019MNRAS.483.3382B}, suggesting diverse orientations or evolutionary stages.

Multiwavelength studies reinforce their jet-linked nature. Flat-spectrum RL-NLSy1s tend to be core-dominated, while steep-spectrum ones often show extended jets \citep{2018A&A...614A..87B}. Broadband SEDs of $\gamma$-NLSy1s resemble those of blazars \citep{2019ApJ...872..169P}, though jet power appears lower than accretion power in some cases \citep{2019ApJ...879..107F}. Jet production in these sources may follow hybrid mechanisms \citep{2022MNRAS.517.1381C}, and mid-IR variability on short timescales has been reported \citep{2021ApJS..255...10M}. SED modelling further supports compact and evolving jets \citep{2023MNRAS.523..404L}.

Intra-night optical variability (INOV) has proven to be a sensitive probe of relativistic jet activity. Strong INOV is common in radio-loud AGN, particularly blazars and high-polarisation quasars, with duty cycles reaching 60–70\% \citep{1995ApJ...452..582J,1998ApJ...501...69D,2005MNRAS.356..607S,2024MNRAS.529L.108O}, but is much rarer in radio-quiet sources. This supports the view that INOV is jet-driven, especially when the jet is closely aligned to the line of sight. In contrast, weaker INOV in radio-quiet quasars likely arises from disc fluctuations or misaligned jets.

\citet{Ojha2022} (hereafter OV22) showed that RL-NLSy1s with $\gamma$-ray detections exhibit high INOV duty cycles, whereas non-jetted RL-NLSy1s show none, highlighting the role of jet orientation and Doppler boosting. This aligns RL-NLSy1s with blazars in their INOV behaviour. However, INOV studies are inherently short-duration lasting typically 3–4 hours, and do not provide a direct empirical test of whether such variability extends to longer timescales.

 While INOV observations have long been described as short {\it snapshots} into jet‑dominated AGN activity \citep{2018BSRSL..87..281G}, the multiyear light curves, give us a much longer {\it snapshot} of AGN variability. This extended baseline allows us to probe whether rapid, jet-induced microvariability also manifests as secular trends over years and whether they scale coherently across timescales. Recent work suggests that variability on month-to-year timescales correlates more closely with accretion-related parameters—such as black hole mass and Eddington ratio—than with jet indicators alone \citep{rakshit2017ApJ...842...96R, 2021Sci...373..789B, 2022MNRAS.517.1381C,2023MNRAS.523..441Y,2025A&A...698A.160R}. This raises the question: do jet-linked fluctuations like INOV translate into coherent, long-term variability?

To test this, we analyse multi-year light curves from the Zwicky Transient Facility (ZTF) \citep{Bellm2019} for a well-studied set of RL-NLSy1s from OV22. Our sample spans three categories—(i) $\gamma$-ray detected, (ii) jetted but $\gamma$-ray undetected, and (iii) non-jetted RL-NLSy1s. We also include a control sample of high-polarisation core-dominated quasars (HPQ), a classical jet-dominated AGN group. Using fractional variability and structure function (SF) diagnostics, we assess whether INOV-active sources also show coherent long-term variability.

An additional dimension in our analysis involves colour–magnitude behaviour, specifically redder-when-brighter (RWB) and bluer-when-brighter (BWB) trends. RWB patterns often signal synchrotron jet emission, while BWB behaviour is associated with thermal disc variability \citep{2011A&A...528A..95G,2015RAA....15.1784Z,2022MNRAS.510.1791N,2024MNRAS.528.4702M}. When interpreted alongside SF evolution and variability amplitudes, these trends help trace the dominant emission mechanism.

The rest of this Letter is organised as follows. Section~2 describes our sample and data. Section~3 presents the variability analysis. Section~4 discusses the results, and Section~5 summarises our conclusions.

\begin{figure*}
\centering

\subfigure{\includegraphics[width=8.5cm,height=5cm]{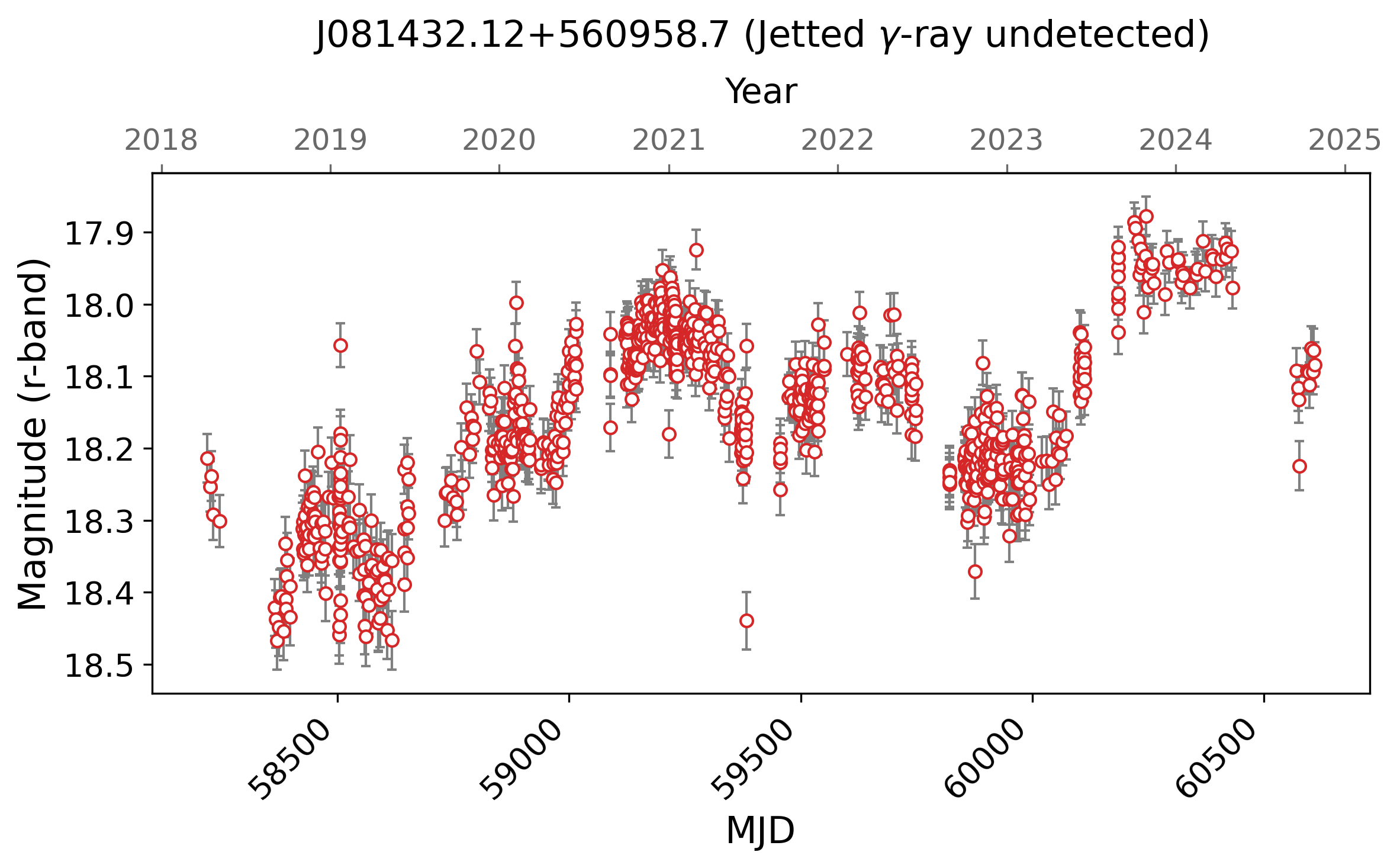}}
\subfigure{\includegraphics[width=8.5cm,height=5cm]{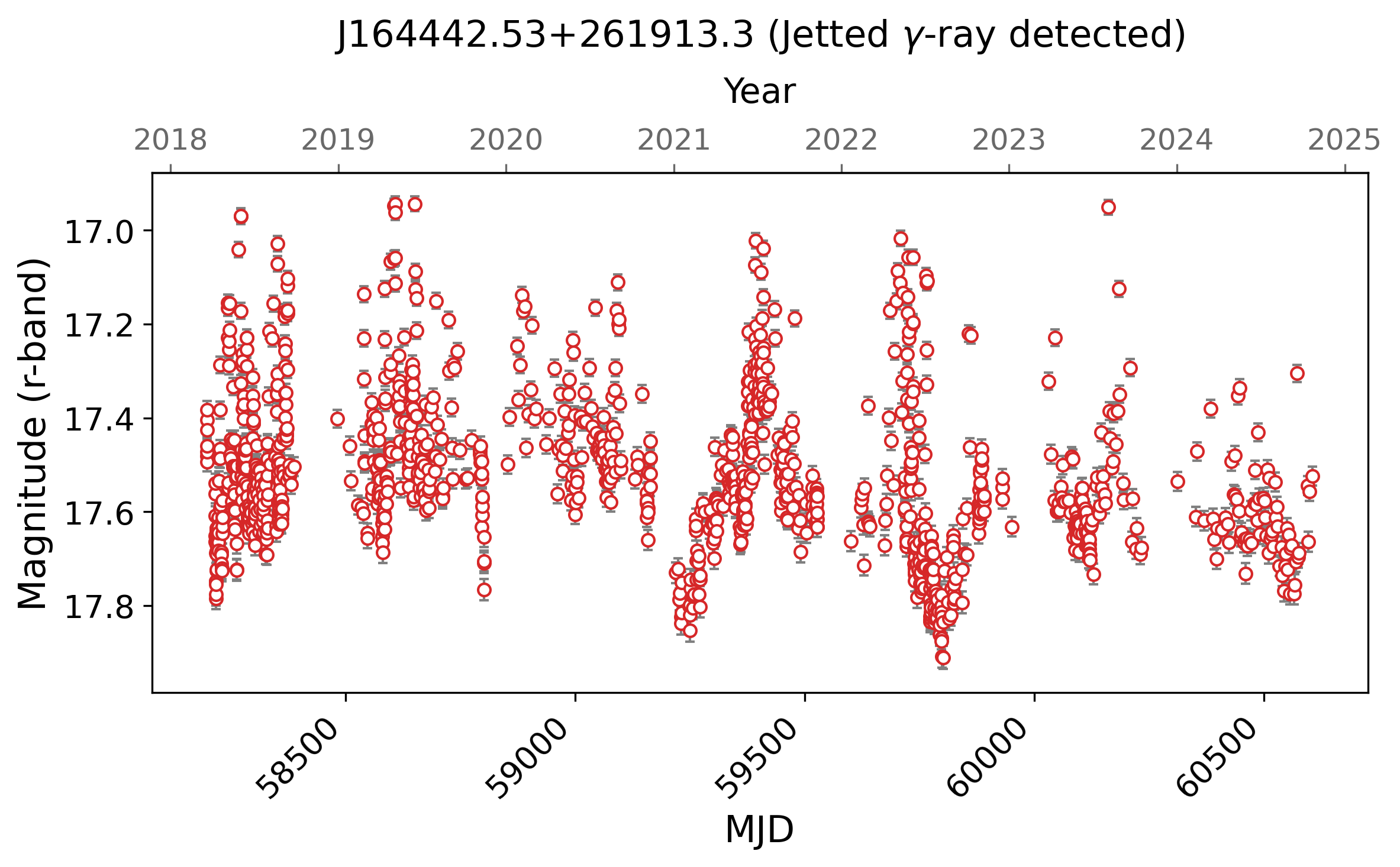}}
\subfigure{\includegraphics[width=8.5cm,height=5cm]{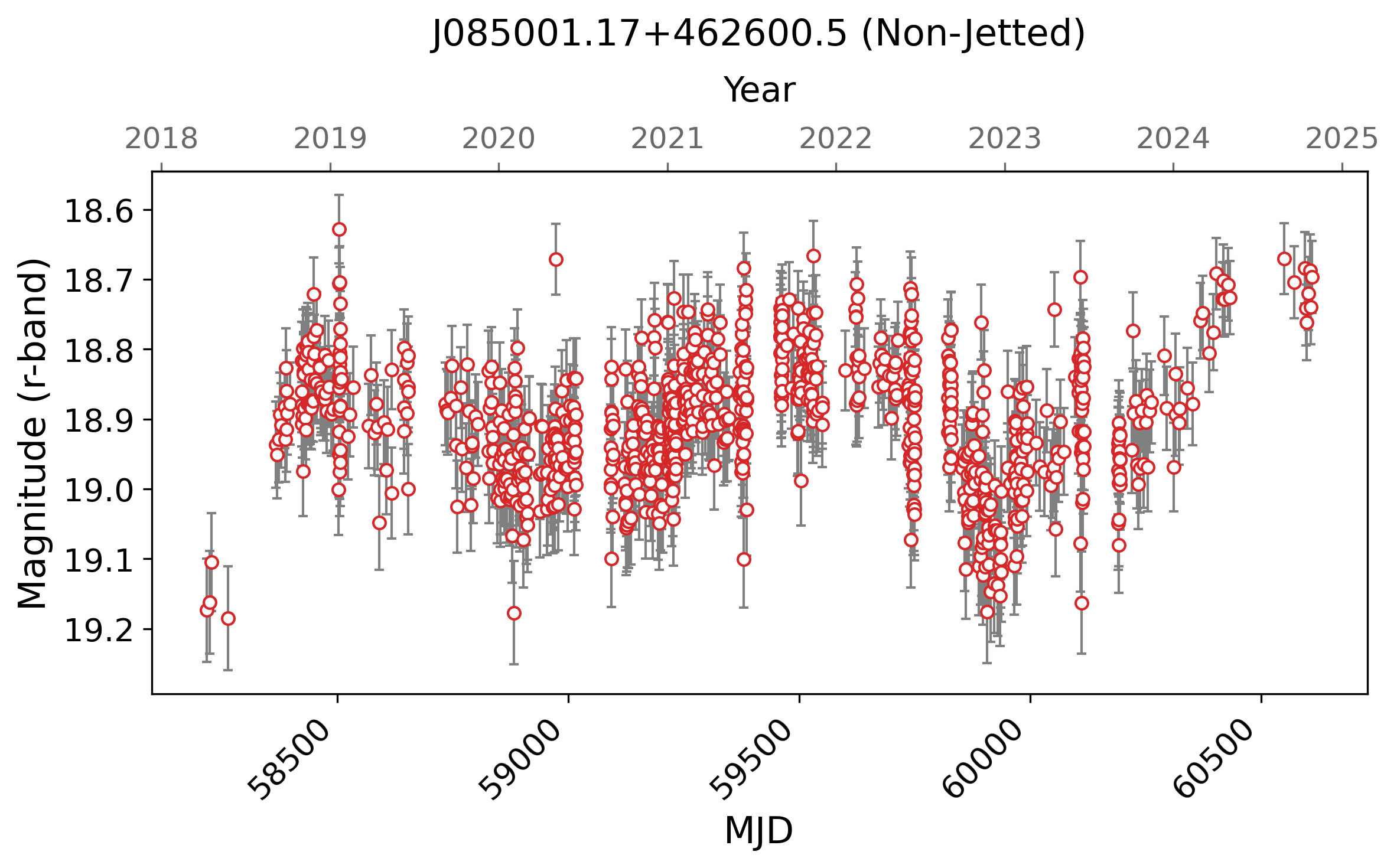}}
\subfigure{\includegraphics[width=8.5cm,height=5cm]{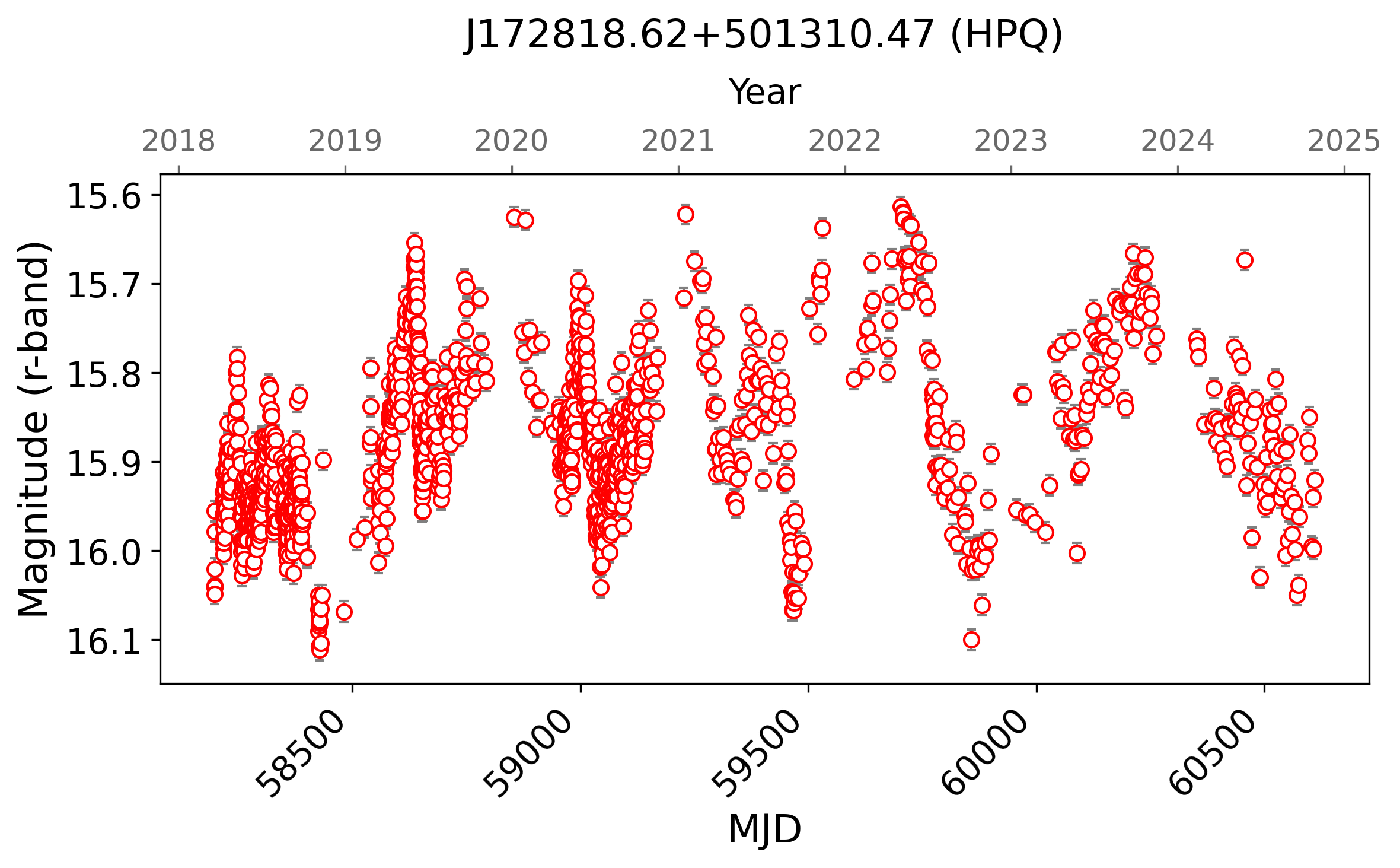}}
% \subfigure{\includegraphics[width=5cm,height=3cm]{lc5.png}}
% \subfigure{\includegraphics[width=5cm,height=3cm]{lc3.png}}

\caption{The ZTF r-band light curves for 4 representative sources studied in this work. The SDSS name for the source, along with their subclass, are noted on the top of each panel. The light curves of all the sources are available as supplementary material. }
\label{light_curves}
\end{figure*}

\section{Sample and Data}

Our primary sample comprises 23 radio-loud Narrow-Line Seyfert 1 (RL-NLSy1) galaxies previously analysed for intra-night optical variability (INOV) by OV22. These sources were selected based on their optical spectral classification, radio loudness (
$R_{1.4\,\mathrm{GHz}} > 100$), and the availability of multi-epoch photometric monitoring from ground-based telescopes. All sources have spectroscopically confirmed redshifts and are grouped into three categories: jetted RL-NLSy1s with confirmed $\gamma$-ray emission detected by \textit{Fermi}-LAT (8 sources); jetted RL-NLSy1s without $\gamma$-ray detection but with VLBI jet signatures (7 sources); and non-jetted RL-NLSy1s lacking both VLBI jet evidence and $\gamma$-ray emission (8 sources).
 We note that the distinction between jetted and non-jetted RL-NLSy1s is based solely on current $\gamma$-ray and VLBI detection limits. Deeper observations may reveal jet signatures in some of the sources presently classified as non-jetted; therefore this subclassification should be regarded as observational rather than intrinsic.

To place the variability of RL-NLSy1s in broader context, we also include a comparison sample of 59 HPQs, and these are radio-loud, mainly powered by synchrotron radiation in radio to optical range, boosted by effects of relativistic motion \citep{2023PASA...40....6C}. This control group serves as a benchmark to assess whether RL-NLSy1s, despite their lower black hole masses and possible younger age, display jet-linked long-term variability properties comparable to classical blazar-type AGN.

We retrieved long-term optical light curves for all sources from the ZTF Data Release 23 (DR23), in both $g$- and $r$-bands via the IRSA portal\footnote{\url{https://irsa.ipac.caltech.edu/Missions/ztf.html}}. The ZTF survey offers high-cadence, wide-field monitoring with a temporal baseline extending from March 2018 to January 2025 ($\sim$ 83 months), enabling variability characterisation on timescales ranging from days to several years (see Figure \ref{light_curves}).

To ensure consistency with the intra-night study by OV22, we focused on $r$-band light curves. We required each light curve to contain at least 100 valid observational epochs in the $r$-band to guarantee adequate temporal coverage and statistical robustness. Additionally we applied the condition \texttt{catflag=0} to confirm that the data contains no known artefacts \citep{Sudan_Madhu10.1093/mnras/staf1441}.  Applying this threshold excluded five sources from RL-NLSy1s sample and seven from HPQs, yielding a final sample of 18 RL-NLSy1 galaxies which included 5 Jetted $\gamma-$ ray undetected,  7 Jetted $\gamma-$ ray detected, 6 Non-jetted  and 52 HPQs for analysis. Additionally, we applied an iterative $3\sigma$-clipping procedure to each light curve to eliminate outliers and spurious measurements that could bias the variability analysis.

\section{Analysis}
\label{sec:analysis}

To quantify long-term optical variability in our sample, we computed two complementary time-series statistics: the fractional variability amplitude ($F_{var}$) and the first-order structure function (SF). These metrics jointly characterise the amplitude and temporal behaviour of variability, allowing us to assess both the total flux dispersion and how variability evolves with timescale.

\subsection{Fractional Variability}
 For each light curve, following \citet{2002ApJ...568..610E} and \citet{Vaughan2003}  we calculated the fractional flux variability amplitude ($F_{var}$), to quantify the degree of variability in a light curve :

\textbf{\begin{equation}
         F_{var} = \sqrt{\frac{S^{2} - \epsilon^{2}}{\overline{m^{2}}}}~,
     \label{eq:fvar_equation}
 \end{equation}}
where,
\begin{equation}
S  = \sqrt{\frac{1}{N-1} \sum_{i=1}^{N}\left(m_{i} - \overline{m}\right)^2}~~.
\end{equation}

In this formulation, the overline (e.g., $\overline{m}$) denotes an unweighted arithmetic mean. $F_{\mathrm{var}}$ quantifies flux variations relative to the mean, corrected for statistical noise, and expressed as a dimensionless, percentage-like statistic. This normalisation enables direct comparison across sources, wavebands, or epochs, unlike raw variance which scales with absolute flux. While easy to compute, $F_{\mathrm{var}}$ is nonetheless affected by the stochastic nature of AGN variability.   The uncertainity in $F_{\mathrm{var}}$ is calculated as: 

 \begin{equation}
 \epsilon (F_{var}) = \sqrt{\left( \sqrt{\frac{1}{2N} }\frac{\epsilon^{2}}{\overline{m^{2}} F_{var}} \right)^{2} + \left( \sqrt{ \frac{\epsilon^{2}}{N} }\frac{1} {\overline{m\
}}  \right)^{2}} ~~
    \label{eq:fvarerr_equation}
\end{equation}

\begin{figure}
    \centering
    \includegraphics[width=\linewidth]{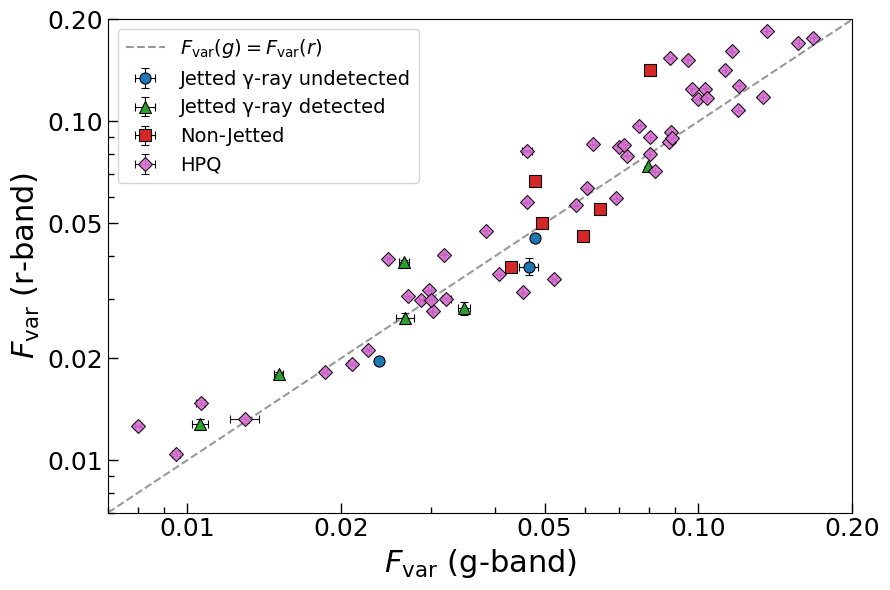}
    \caption{
Comparison of fractional variability amplitude ($F_{\mathrm{var}}$) in the $ g$- and $r$-bands for different AGN classes. Blue circles represent jetted RL-NLSy1 sources without $\gamma$-ray detection, green triangles denote $\gamma$-ray detected jetted sources, and red squares correspond to non-jetted NLSy1, while pink diamonds represent the HPQs.  The dotted line with 45$^{\circ}$ slope refering to $F_{var(g)}=F_{var(r)}$ is drawn to guide the eye.}

    \label{fig:fvar}
\end{figure}

In the above formulation, the overline notation (e.g., $\overline{m}$) denotes a simple arithmetic mean, i.e., without any weight. The variability amplitude was computed separately for the $ g$- and $r$-band light curves, allowing a direct comparison of variability amplitude across filters (see Figure \ref{fig:fvar} and Table \ref{tab:source-properties}).

\subsection{Structure Function}

The structure function (SF) is a standard tool for characterising variability timescales and periodicities \citep[e.g.][]{2002MNRAS.329...76H,2005AJ....129..615D,2011A&A...525A..37M,graham2014novel}. It can be computed for individual light curves or as an ensemble SF from multiple sources \citep[e.g.][]{Ritish2025,Sudan_Madhu10.1093/mnras/staf1441}, representing the rms magnitude difference versus time lag. Binning measurements makes SF analysis well suited to large datasets, with bins being statistically independent \citep[e.g.][]{2005AJ....129..615D}.  We measure variability strength as a function of timescale using a modified first-order SF \citep{1996ApJ...463..466D}:

\begin{equation}
SF_{(\Delta t)} = \sqrt{\frac{\pi}{2}\langle |m(t+\Delta t)-m(t)|\rangle^2 - \langle \sigma_{n}^2 \rangle}~,
\end{equation}

where $m(t)$ is the magnitude and $\langle \rangle$ denotes averages over the light curve. We use logarithmic lag bins (0.2 dex) with at least 10 pairs per bin for robustness. Squaring after averaging in the first term reduces outlier impact \citep{1994MNRAS.268..305H}. The noise term $\sigma_{n}$ is the standard deviation of each pair separated by $\Delta t$. The factor  $\pi/2$ assumes Gaussian intrinsic variability and noise \citep{wilhite2008variability}.

\begin{table*}
\caption{Summary of the RL-NLSy1 and comparison HPQ samples used in
this study. The table presents key properties for each source, with variability
metrics derived from ZTF r-band light curves. The INOV information for the Jetted and non jetted sources is available from OV22, while for the HPQ the INOV information is taken from   \citep{Goyalarti}. The entire table for all the sources is available online as suplementary material.}
\label{tab:source-properties}
\centering
\footnotesize
\begin{tabular}{lcccccc}
\hline
\textbf{SDSS Name} & \textbf{Subclass} & \textbf{Redshift} & \textbf{$F_{\mathrm{var}}$ (r-band)} & \textbf{\boldmath $<\theta>^a$}
& \textbf{Colour Trend$^b$} %\tablefootnote{\scriptsize BWB: Bluer When Brighter, RWB: Redded When Brighter}} 
& \textbf{INOV status$^c$}
%\tablefootnote{\scriptsize V: Variable, PV: Partially Variable, NV: Non Variable}}
\\
\hline
J081432.12+560958.7 & Jetted $\gamma-$ ray undetected& 0.51 & 0.020$\pm 0.0004$ &37.33 $\pm 0.41$ &BWB & NV \\
J085001.17+462600.5 & Non-jetted      & 0.52 & 0.142 $\pm 0.0012$  & 43.39 $\pm0.75$&BWB & NV \\
J164442.53+261913.3 & Jetted $\gamma-$ ray detected   & 0.14 & 0.026 $\pm 0.00049$ & 47.19 $\pm0.43$&RWB & PV \\

J172818.62+501310.4 & HPQ & 0.05 & 0.019$\pm 0.0002$ & 47.00 $\pm0.30$&RWB & $-$ \\

--- & --- & --- & --- & --- & --- \\
\hline
\multicolumn{7}{l}{\textbf{\boldmath $<\theta>^a$}: Mean $\theta$.}\\
\multicolumn{7}{l}{$^b$BWB: Bluer When Brighter, RWB: Redder When Brighter.}\\
%\multicolumn{6}{l}{\textbf{$^b$V: Variable, PV: Probable Variable, NV: Non Variable. }}\\
\multicolumn{7}{l}{$^c$V=variable , i.e. confidence level $\geq$ 99\%;PV = probable variable, i.e. $95-99$\% confidence level; NV = non-variable, i.e. confidence level $<$ 95\%.}\\

\end{tabular}

\end{table*}

Structure functions were calculated in rest frames for each source. To compare ensemble behaviour across physical classes, we constructed mean SF profiles for all the four subgroups (jetted with $\gamma$-ray, jetted without $\gamma$-ray, non-jetted and HPQ sources), using a robust biweight mean to mitigate the effect of outliers.

\section{Results and Discussion}

\subsection{Variability Amplitudes}

Figure ~\ref{fig:fvar} shows the fractional variability amplitudes ($F_{\rm var}$) in the $g$ and $r$ bands for the 18 RL-NLSy1 galaxies in our sample and the 52 HPQs. The lowest long-term variability, as measured by $F_{\rm var}$, is observed in the Jetted $\gamma$-ray undetected RL-NLSy1s, with the median value of 0.02.  Sources that are jetted and $\gamma$-ray detected have a slightly higher $F_{\rm var}$ with the median being 0.026. Their $F_{\rm var}$ values are tightly clustered in the range 0.01--0.02, a finding that contrasts with their high intra-night optical variability (INOV) as reported by OV22, with the Duty Cycle (DC) of 34 \%. Conversely, the non-jetted RL-NLSy1s exhibit the highest $F_{\rm var}$ values, occasionally exceeding 0.05, while OV22 report their INOV duty cycle as 5\%. This suggests that the long-term variability amplitude does not directly scale with intra-night variability behaviour. In contrast to the clear separation observed among the RL-NLSy1 subclasses, our HPQ comparison sample spans the entire range of $F_{\rm var}$ values (see Table \ref{tab:sample} for median values of each class). This wide range indicates that, for this class of objects, $F_{\rm var}$ is not a simple discriminator, a result consistent with the known broad distribution of variability amplitudes in such sources \citep{2022MNRAS.510.1809P}.

\subsection{Structure Function Evolution with Time}

\begin{figure*}
    \centering
    \subfigure{\includegraphics[width=8.5cm,height=5.2cm]{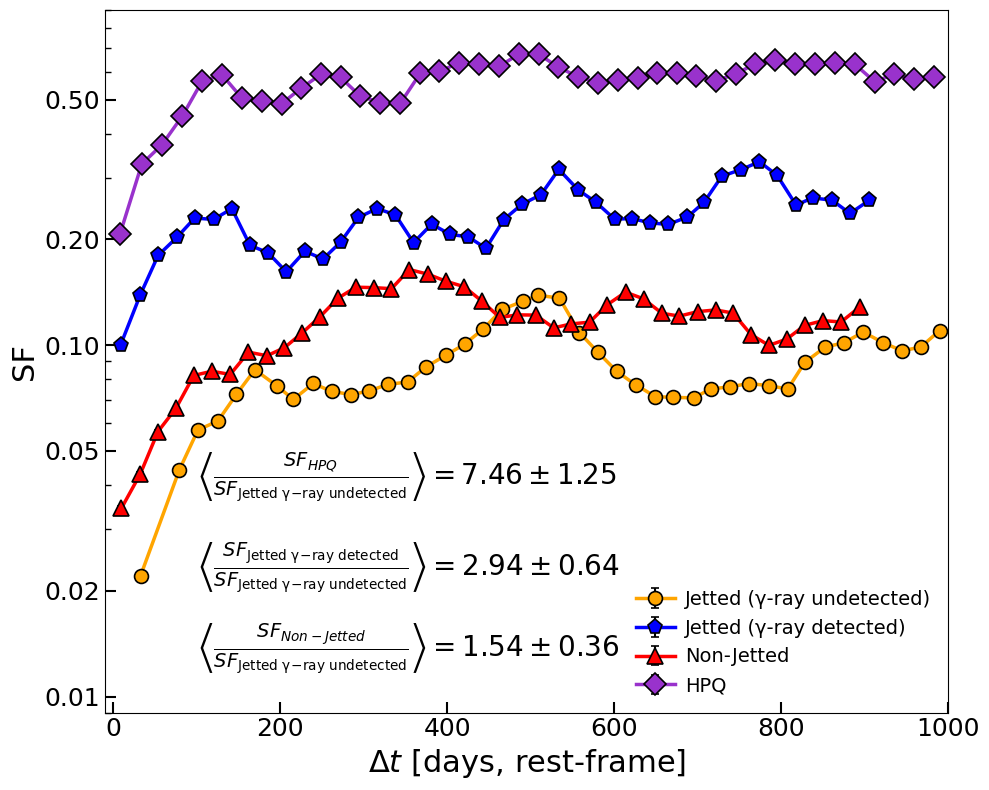}}
    \subfigure{\includegraphics[width=8.5cm,height=5.2cm]{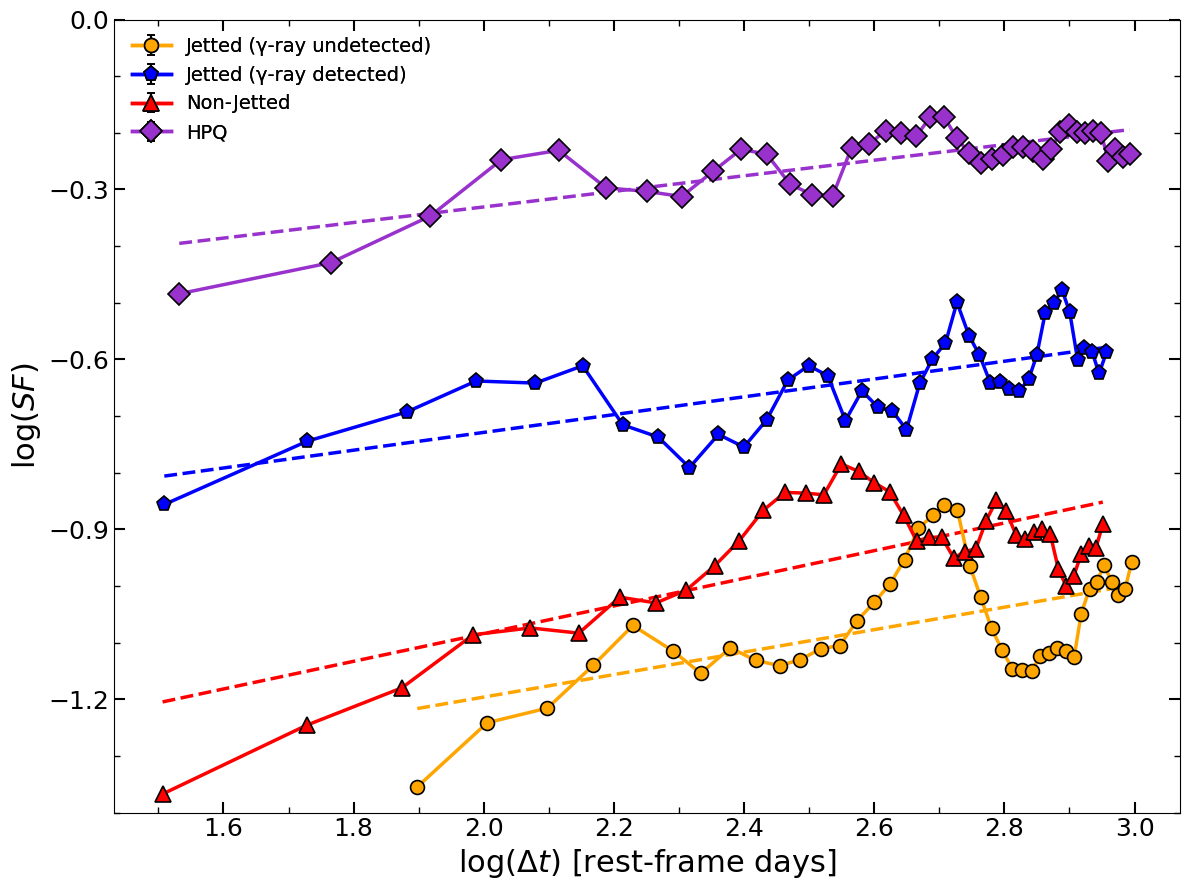}}
    
    \caption{
\emph{Left:} Structure function (SF) comparison including the three NLSy1 subclasses of $\gamma$-ray detected RL-NLSy1s (blue pentagons), $\gamma$-ray undetected RL-NLSy1s (orange circles), and non-jetted NLSy1s (red triangles), and the highly polarised core-dominated quasars (HPQ) as an additional jet-dominated class (purple diamonds). Both the $\gamma$-RL-NLSy1 and HPQ groups show rising SFs with higher variability amplitude and long-term coherence, while the other two classes plateau after a certain duration.  \emph{Right:} This panel shows the ensembled SF in log–log space with best-fitting power-law slopes (SF $\propto \Delta t^\beta$). The best-fit slopes for different subclasses are listed in the second-to-last column of Table~\ref{tab:sample}. Error bars in all plots are smaller than the symbol sizes and are therefore not visible.}

    \label{fig:sf_hpq}
\end{figure*}

To investigate variability as a function of timescale, we computed rest-frame structure functions (SFs) for each source. In Figure \ref{fig:sf_hpq}, we plot the SF including our sample of HPQs. It is clearly seen that the HPQs exhibit the steepest and most elevated structure function across all timescales, followed closely by the $\gamma$-ray detected jetted RL-NLSy1s, while the non-jetted and $\gamma$-ray undetected sources show flatter SFs with lower amplitudes, indicating a lack of long-term coherence in their variability. A clear separation emerges, with the $\gamma$-ray detected jetted RL-NLSy1s exhibiting a steadily rising SF up to the sampled timescales ($\sim$1000 days), indicating coherent long-term variability. In contrast, non-jetted sources show relatively high SF at short lags ($\lesssim 500$ days), but the SF plateaus or declines at longer lags. This plateauing behaviour is a hallmark of a Damped Random Walk (DRW) process, suggesting that the variability of these disk-dominated sources is governed by a characteristic timescale, $\tau$ \citep{2016ApJ...826..118K}. Meanwhile, the jetted but $\gamma$-ray undetected sources show a modest rise and plateau, but do not reach the levels seen in their $\gamma$-loud counterparts.  The uncertainties in the structure function are included in the plot, but are smaller than the plotted symbols and thus not readily visible.

We quantify the class differences by computing the mean SF amplitude at $\Delta t = 1000$ days. The ratio between $\gamma$-ray detected jetted and $\gamma$-ray undetected jetted SFs is $\langle {\rm SF}_{\rm jetted\ \gamma\text{-ray\ detected}} / {\rm SF}_{\rm jetted\ \gamma\text{-ray\ undetected}} \rangle = 2.94 \pm 0.64$, while the ratio relative to non-jetted sources is $\langle {\rm SF}_{\rm Non\text{-}jetted} / {\rm SF}_{\rm jetted\ \gamma\text{-ray\ undetected}} \rangle = 1.54 \pm 0.36$. These differences are statistically significant and reflect a strong link between jet orientation, $\gamma$-ray emission, and long-timescale variability coherence.

In addition, we now present the ensemble structure functions in log–log space, showing the best-fitting power-law slopes, following the approach of \citet{2018MNRAS.478.2557G}. The derived slopes are $\beta$ = 0.20 $\pm$ 0.05 for jetted $\gamma$-ray undetected, 0.24 $\pm$ 0.04 for non-jetted RL-NLSy1s, 0.16 $\pm$ 0.03 for jetted $\gamma$-ray detected, and 0.14 $\pm$ 0.02 for HPQs. These slopes are shallower than the canonical DRW value ($\beta$ $\sim$ 0.5) \citep{2016ApJ...826..118K}, suggesting that the optical variability in these sources saturates on timescales shorter than the observational baseline. The modest differences in $\beta$ across subclasses indicate that while the variability amplitude depends strongly on jet strength, the underlying stochastic timescale remains broadly similar.

\begin{table*}
\centering
\scriptsize
\caption{Summary of the median values of properties of RL-NLSy1 and comparison HPQ samples used in this study.}
\begin{tabular}{lrrrrrrr}
\hline
 Subclass & \textbf{N$^a$}
 %\tablefootnote{\scriptsize Number of sources} 
 & INOV DC$^b$
 & Redshift & $F_{var}$ ($r$-band)  & \textbf{SF Slope} & \textbf{$\overline{\theta}$ $^c$} & Dominant Colour\\
 & & (\%) & (median) &(median)   & (degrees)&  &  (BWB/RWB) (\%)\\
\hline
Jetted $\gamma$-ray undetected & 5 &0\% & 0.66 & 0.0196$\pm$ 0.0001  & 0.20 $\pm$ 0.05 & 40.00 $\pm$ 1.21 & BWB (80\%)\\
Non-jetted& 6 &5\% & 0.51 & 0.0526$\pm$ 0.0004  & 0.24 $\pm $0.04 &  41.57 $\pm$ 1.13 & BWB (83\%)\\
Jetted $\gamma$-ray detected & 7 & 34\% & 0.58 & 0.0262$\pm$ 0.0006 &0.16 $\pm$ 0.03 & 42.96 $\pm$ 0.87 & BWB (57\%) \\
HPQ& 52 & 40.5\% & 0.58 & 0.0809$\pm$ 0.0001 & 0.14 $\pm$ 0.02 & 48.82 $\pm$ 0.67 & RWB (71\%) \\
\hline
\multicolumn{8}{l}{$^a$Number of sources.}\\
\multicolumn{8}{l}{$^b$The duty cycle (DC) values for the first three subclasses are adopted from OV22. For the HPQ subclass, we employ the measurements reported  }\\
 \multicolumn{8}{l}{by \citet{Goyalarti}, who determined DCs using two independent methodologies and further subdivided their sample according to the peak-to-peak }\\
 \multicolumn{8}{l}INOV amplitude. For consistency,  we use the arithmetic mean of all four reported values as the representative DC for our analysis.\\ 
\multicolumn{8}{l}{$^c$ This column shows the median value of the average $\theta$ computed for each source using Eq. 5,  for each subclass with its associated median error.} \\
 \multicolumn{8}{l}{The value of $\overline{\theta}$ below $45^\circ$ represents BWB trend and above $45^\circ$ it represent RWB behaviour. This threshold of $45^\circ$  marks the reference case $\Delta f_{r} = \Delta f_{g}$; }\\
 \multicolumn{8}{l}{ points below (above) this threshold shows the relative dominance of bluer (redder) flux.} 
\end{tabular}
\label{tab:sample}
\end{table*}

\subsection{A Disconnect Between Amplitude and Temporal Coherence} 

\noindent The disparity between variability metrics is central to our results. $\gamma$-ray detected jetted RL-NLSy1s show low integrated variability amplitudes ($F_{\rm var}$) yet rising structure functions, indicating coherent long-term trends. In contrast, non-jetted sources have high $F_{\rm var}$ but plateauing SFs, consistent with stochastic, uncorrelated variability on shorter timescales and the DRW model. This disconnect shows that amplitude alone does not capture AGN variability. The highly-polarised core-dominated quasars (HPQs) reinforce this: they exhibit the highest SF amplitudes, followed by $\gamma$-ray detected RL-NLSy1s, while $\gamma$-ray undetected and non-jetted sources remain flatter, lacking long-term coherence. Putting our long-term variability results in the context of the study by OV22, we note that the INOV DC reaches $\sim$34\% for $\gamma$-ray detected jetted RL-NLSy1s, but is essentially zero for the non-jetted class. By contrast, our $F_{\rm var}$ measurements show the inverse trend: the $\gamma$-ray detected jetted sources exhibit the lowest long-term variability amplitudes, while the non-jetted systems occupy the upper envelope. This indicates that INOV duty cycles, which capture short-timescale jet activity, do not scale directly into long-term optical variability trends.

This supports a model where powerful jets establish variability coherence via Doppler boosting or slow jet-structural changes \citep{2007A&A...469..899H}, while disc-dominated systems produce bursty, incoherent variability from local instabilities \citep{1997ARA&A..35..445U, 2017A&ARv..25....2P}. Our findings agree with blazar studies showing rising SFs from coherent variability \citep{2009ApJ...699.1732B}, but reveal that the effect is not universal to all jetted AGNs. Our $\gamma$-ray detected RL-NLSy1s considered likely strongly beamed analogues, maintain low $F_{\rm var}$ despite jet dominance, indicating that temporal coherence from the SF is a better jet diagnostic than amplitude. This underscores the role of jet orientation and relativistic effects in shaping long-term optical variability.

\subsection{Colour–Magnitude Behaviour}

Active galactic nuclei (AGN) show colour variability linked to flux changes. We constrain this behaviour over months–years using quasi-simultaneous $\mathrm{g}$- and $\mathrm{r}$-band observations within $\sim$30-min gaps. Earlier studies measured colour variability via colour–magnitude fits (e.g.,$g$ vs.\ $(g-r)$; \citealt{8135723, 2004ApJ...601..692V, 2005ApJ...633..638W}), but this approach is affected by colour–magnitude error covariances. \citet{2012ApJ...744..147S} mitigated this by fitting in magnitude–magnitude space and translating results to colour–magnitude space. Literature \citep[e.g.][]{2015A&A...581A..93R,Sudan_Madhu10.1093/mnras/staf1441} shows that, in magnitude–magnitude space, observed magnitudes combine variable AGN emission and constant host flux, yielding a nonlinear relation. The common {\it bluer when brighter} (BWB) trend in total flux can thus result from a red, constant host (including non-varying lines) plus a variable blue AGN continuum \citep[e.g.][]{Sakata2010ApJ...711..461S, 2015A&A...581A..93R}. To correct for this, we converted $\mathrm{g}$- and $\mathrm{r}$-band magnitudes to fluxes using the zero-points from \citet{1999AJ....118.1406L}\footnote{\url{https://www.sdss4.org/dr17/algorithms/magnitudes/}}. We then calculated the colour variation angle in flux–flux space, as defined by \citet{2015A&A...581A..93R}:

\begin{equation}
\theta = \arctan\left( \frac{f_r(t + \tau) - f_r(t)}{f_g(t + \tau) - f_g(t)} \right),
\label{theta_flux}
\end{equation}
where $f_g$ and $f_r$ are the $\mathrm{g}$- and $\mathrm{r}$-band fluxes, respectively. The average $\theta$ for each source was determined by averaging over all valid epoch pairs in its light curve. 

The median values of average $\theta$ for each class, alongwith associated error, are given in the seventh column of Table~\ref{tab:sample}.

The most jet-dominated sources, the highly-polarised quasars (HPQs), show a clear redder-when-brighter (RWB) trend (71\% of sources) , indicative of a steepening non-thermal spectrum from variable synchrotron emission during high states, as commonly seen in blazars and other jet-aligned systems \citep{2011A&A...528A..95G,2022MNRAS.510.1791N}. Non-jetted RL-NLSy1s instead favour a bluer-when-brighter (BWB) trend (83\% of the sample)or no correlation, consistent with thermal emission from accretion-disc fluctuations \citep{2006A&A...450...39G}. The $\gamma$-ray undetected RL-NLSy1s also show a dominant BWB trend with 80\% of sources. Interestingly, $\gamma$-ray detected RL-NLSy1s show mixed behaviour (57\% sources with BWB trend and 43\% sources with RWB trend), suggesting a transitional class where optical output blends variable, orientation-dependent jet emission with disc fluctuations. These results support a two-component model in which jet–disc interplay shapes colour–variability patterns \citep{2010MNRAS.404.1992R}.  Although the present analysis suggests mild bluer-when-brighter behaviour in the jetted RL-NLSy1s, the small number of sources with reliable colour information prevents a firm statistical conclusion. We therefore regard these results as indicative, and note that a larger sample will be needed for a more robust characterisation of colour-variability trends across AGN subclasses.

A recent study by \citet{2025arXiv250503902O} attributes variability in non–$\gamma$-ray-detected NLSy1s to accretion-disc instabilities. This aligns with our results for disc-dominated systems but underscores a key point: amplitude measures such as $F_{\rm var}$ alone cannot reveal the variability driver. By using the structure function (SF) to probe coherence, we find the RWB trend in $\gamma$-ray detected sources is best explained by a coherent, jet-driven component. Timescale-sensitive diagnostics like the SF thus capture long-term variability often missed by amplitude-based analysis, offering a fuller view of the jet–disc connection.

\section*{Conclusions}
We investigated the optical variability of a well-characterised sample of radio-loud NLSy1 galaxies (both $\gamma$-ray detected and undetected), together with highly polarised core-dominated quasars (HPQs) and non-jetted, using ZTF light curves and historical INOV data. Our key results are:
\begin{itemize}
    \item The fractional variability amplitude ($F_{\rm var}$) is lowest in $\gamma$-ray undetected RL-NLSy1s (median value of 0.020), intermediate in $\gamma$-ray detected jetted RL-NLSy1s (median value of 0.026), and highest in non-jetted RL-NLSy1s (median value of 0.052), showing that jet presence does not guarantee larger multi-year variability amplitudes.

    \item Structure function (SF) analysis indicates clear stratification in temporal coherence: HPQs show the steepest and highest SFs, followed by $\gamma$-ray detected RL-NLSy1s, $\gamma$-ray undetected RL-NLSy1s, and Non-jetted RL-NLSy1s. $\gamma$-ray detected RL-NLSy1s display monotonic SF growth over $\sim$1000\,d, consistent with a persistent jet-linked component, whereas non-jetted sources saturate at shorter timescales, characteristic of stochastic disc variability.

    \item The divergence between $F_{\rm var}$ and SF results demonstrates that amplitude-based metrics alone can misidentify variability drivers; timescale-sensitive diagnostics such as the SF are essential for detecting coherent, jet-induced variability.

    \item Colour--magnitude trends further separate the populations: HPQs predominantly show a redder-when-brighter (RWB) trend from synchrotron emission, Non-jetted RL-NLSy1s a bluer-when-brighter (BWB) trend from disc variability, and $\gamma$-ray detected RL-NLSy1s a mixture, consistent with a two-component disc–jet model.
\end{itemize}

This study shows that variability across timescales is a powerful diagnostic of both the presence and dominance of relativistic jets. By integrating multi-timescale variability (via structure functions) with colour behaviour, we can more fully characterise the jet–disc connection in AGNs. Our framework is well suited to future large-scale time-domain surveys such as the Legacy Survey of Space and Time (LSST) which will enable systematic identification of jetted AGNs and tracking of jet-induced variability in thousands of RL-NLSy1s, blazars, and related systems. Ultimately, a combined short- and long-term approach is essential to fully investigate the physics of jet–disc coupling.

\section*{Acknowledgements}

We thank the anonymous referee for the helpful comments which substantially improved the manuscript. H.C. and A.K.S. sincerely thank the Inter-University Centre for Astronomy and Astrophysics (IUCAA) for their hospitality and for providing access to their High Performance Computing (HPC) facilities through the IUCAA Associate Programme. M.S. would like to thank the University Grants Commission (UGC), Government of India, for support under the UGC–JRF/SRF scheme (Ref. No.: 221610066226). Based on observations obtained with the Samuel Oschin Telescope 48-inch and the 60-inch Telescope at the Palomar Observatory as part of the Zwicky Transient Facility project. ZTF is supported by the National Science Foundation under Grants No. AST-1440341 and AST-2034437 and a collaboration including current partners Caltech, IPAC, the Oskar Klein Center at Stockholm University, the University of Maryland, University of California, Berkeley , the University of Wisconsin at Milwaukee, University of Warwick, Ruhr University, Cornell University, Northwestern University and Drexel University. Operations are conducted by COO, IPAC, and UW. 
%%%%%%%%%%%%%%%%%%%%%%%%%%%%%%%%%%%%%%%%%%%%%%%%%%
\section*{Data Availability}

The data used in this study are publicly available in the Zwicky Transient Facility (https://www.ztf.caltech.edu/) data release 23. Derived parameters and related fits are available upon request.



\bibliographystyle{mnras}
\bibliography{main} 
%%%%%%%%%%%%%%%%%%%%%%%%%%%%%%%%%%%%%%%%%%%%%%%%%%

%%%%%%%%%%%%%%%%% APPENDICES %%%%%%%%%%%%%%%%%%%%%

%%%%%%%%%%%%%%%%%%%%%%%%%%%%%%%%%%%%%%%%%%%%%%%%%%

% Don't change these lines
\bsp	% typesetting comment
\label{lastpage}
\end{document}